# Thickness-dependent properties of transparent conductive oxides for epsilon-near-zero applications.

J. Gosciniak[1*]

[1]*Institute of Microelectronics and Optoelectronics, Warsaw University of Technology, Koszykowa 75, 00-662 Warsaw, Poland*
*Corresponding authors: * jacek.gosciniak@pw.edu.pl*

**Abstract**
In recent years, epsilon-near-zero (ENZ) materials have attracted much attention due to their unique properties that can be tuned under electrical and optical signals. Furthermore, they allow for a strong enhancement of a nonlinearity close to the ENZ regime, which can have a direct impact on many fields ranging from telecommunications, sensing, quantum optics, to neuromorphic computing. Among the many materials belonging to the ENZ class, transparent conductive oxides (TCOs) are of particular interest because they are very well known to the community and possess very well understood electrical and optical properties. This work extends the direct control of the optical properties of transparent conductive oxides by tailoring the film thickness, which opens new possibilities for ENZ-enhanced photonic applications. It is shown that the thickness-dependent ENZ resonance in the TCO films is equivalent to the power-dependent ENZ resonance as for the same amount of power provided to the TCO film the electric field confinement in the thinner films is highly enhanced for TM polarized light, resulting in the higher electrons effective mass. Thus, the optical properties of TCO can be optically tuned in the ENZ regime, which opens possibilities for a new type of devices that can operate under all-optical switching mechanism.

**Introduction**
With the increasing complexity of the optoelectronic devices, which are often composed of a multiple stack of materials, controlling the optical and electrical response of each of the materials that make up the device is essential to achieving the desired functionality [**1-3**]. One of the solutions is to control the optical and electrical properties during growth by varying the deposition conditions and post-deposition processing. It can be related to different deposition temperature, gas ratio, dopant concentration, post-deposition annealing and etc. [**4**]. Apart from this, the response of the material can be controlled externally by different arrangement of the signal provided to the system (optical or electrical), which can involve a unique interaction of the signal with a material [**4-10**]. This is particularly important for thin materials that exhibit a geometrical anisotropy and an increased surface-to-volume ratio, which determines how much of the material is exposed to the signal. One of the examples of such materials are transparent conductive oxides (TCOs), where the epsilon-near-zero (ENZ) resonances can be controlled either by growth conditions, post-deposition annealing, an external signal or even a polarization of light, and thus, the coupling arrangement.
Transparent conductive oxides belong to the ENZ materials that have proven to be excellent tunable materials due to their large permittivity tunability under an applied voltage or a light illumination [**4, 6, 7, 13**]. They are characterized by low optical losses, fast switching time and low switching voltage. In addition, they are CMOS-compatible and can be mass-produced using standard fabrication methods [**4**]. Their unique optical properties for thin materials can be tailored under different polarization of light and coupling arrangement [**6, 7, 11-12, 14**]. They are also characterized by giant ultrafast thermal and optical nonlinearities [**15-20**].
The family of transparent conductive oxides (TCOs) is broad ranging from indium tin oxide (ITO), indium zinc oxide (IZO), cadmium oxide (CdO), gallium zinc oxide (GZO), aluminum zinc oxide (AZO) to mention only a few of the most popular compounds [**4, 21**]. Thus, they can operate in a wide range of wavelengths from UV to MIR. Furthermore, unlike noble metals, the operating wavelength can be tuned either under electrical or optical signals which opens new possibilities in terms of optoelectronic devices with tunable characteristics [**4-13**].

**TCO material properties**

Here we focused on the most popular TCO materials, *i.e.*, ITO [**10, 22**], AZO [**14, 23, 24**], GZO [**25, 26**], and CdO [**8, 9, 27**], which exhibit the ENZ wavelength in a wide spectral range from 1000 nm for GZO up to 3300 nm for highly doped AZO (10 % doped Al). The basic properties of these materials used in the simulation are shown in **Table 1**.

**Table 1**. Optical properties of some common TCO materials.

|  | ITO [10] | AZO [14] | GZO [26] | CdO [26] |
|---|---|---|---|---|
| **High frequency permittivity, $\varepsilon_\infty$** | 3.9 | 3.8825 | 2.475 | 5.5 |
| **Plasma frequency, $\omega_p$ (rad/s)** | $2.52 \cdot 10^{15}$ | $1.14 \cdot 10^{15}$ | $2.93 \cdot 10^{15}$ | $2.41 \cdot 10^{15}$ |
| **Damping factor, $\gamma$ (rad/s)** | $1.80 \cdot 10^{14}$ | $1.27 \cdot 10^{14}$ | $1.78 \cdot 10^{14}$ | $3.06 \cdot 10^{13}$ |

As observed, the damping factor for all TCO materials shown in **Table 1** is about one order of magnitude lower compared to their plasma frequencies. The CdO is an exception as a damping factor is two orders of magnitude lower compared to its plasma frequency as a result of a one order of magnitude higher carrier mobility compared to other TCO materials [**8, 9, 27**].

Based on the parameters presented in **Table 1**, **Fig. 1** was created showing the real and imaginary parts of the permittivity as a function of wavelength (**Fig. 1a**) and plasma frequency (**Fig. 1b**). The real and imaginary parts of the permittivity for TCO materials were calculated using the Drude formula:

$$\varepsilon = \varepsilon_\infty - \frac{\omega_p^2}{\omega^2 - i\omega\gamma} \qquad (1)$$

where $\varepsilon_\infty$ is the permittivity due to bound electrons, $\omega_p$ is the plasma frequency, $\omega$ is the angular frequency of light, and $\gamma$ is the scattering rate, the damping factor.

It can be observed that AZO is characterized by the lowest plasma frequency, while the plasma frequency of GZO is the highest. This is reflected in the ENZ wavelength shown in **Fig. 1a**, where the ENZ wavelength for GZO is the lowest at 1000 nm while the ENZ wavelength for AZO was calculated at 3300 nm.

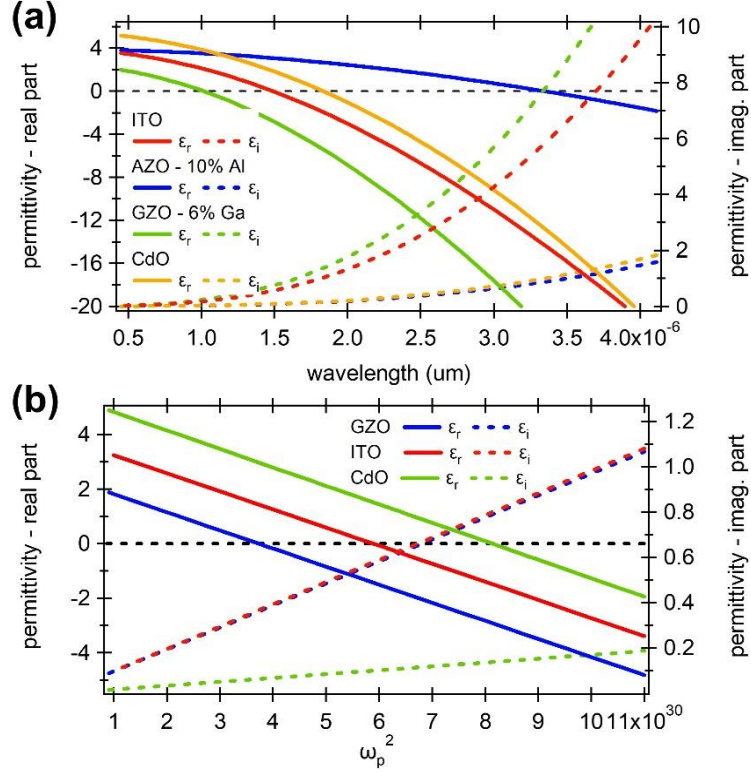

Fig. 1. (a, b) Dispersion of real and imaginary parts of dielectric permittivity of ITO, GZO, AZO and In:CdO as a function of (a) wavelength and (b) plasma frequency.

TCO materials are an important class of materials for optoelectronic applications as they are characterized by the nonparabolic conduction band, thus the energy-wavevector (E-k) dispersion relation of the conduction band is expressed by:

$$\frac{\hbar^2 k^2}{2m^*} = E + CE^2 \qquad (2)$$

where $\hbar$ is the reduced Planck constant, $k$ is the electron wavevector, $m^*$ is the electron effective mass at the conduction band minimum, $E$ is the electron energy at the conduction band minimum, and $C$ is the constant known as the first-order nonparabolicity factor [**19, 20, 28**]. Nonparabolicity is defined here by an extra terms $CE^2$ in the E-k diagram, which introduces a nonlinearity and causes the effective mass $m^*$ to become a function of the electron density $N_c$, i.e., the effective mass increases with increasing carrier concentration under interband absorption of light or an electrical doping. As the carrier density increases, the conduction band fills up and higher band regions, characterized by higher effective masses, become relevant for the optical properties [**28**]. Thus, interband absorption of light causes the carrier concentration to increase, which increases the plasma frequency and blueshifts the ENZ wavelength according to the equation:

$$\omega_p^2 = \frac{N_c e^2}{\varepsilon_0 m^*} \qquad (3)$$

where $N_c$ is the carrier concentration, $e$ is the electric charge, $\varepsilon_0$ is the background permittivity, and $m^*$ is the effective mass. However, as it has been mentioned above, the increased carrier concentration raises the Fermi level what increases the effective mass $m^*$ due to the nonparabolicity. Thus, for materials with high nonparabolicity, the increase in the effective mass can dominate the increase in carrier concentration, and the interband transition can cause an overall redshift of the ENZ wavelength, inverting the nonlinearity [**17**]. Consequently, the overall magnitude and direction of the plasma frequency shift and the nonlinearity change depend on the nonparabolicity of the conduction band and the carrier density.

In addition, the effective mass can be tuned by the intraband absorption of light. Under light absorption, the energy is transferred to the electrons in the TCO, resulting in an increase in the electron temperature $T_e$ and, consequently, the electrons move to higher energy states in the conduction band with higher effective mass $m^*$ [**19, 20**]. At the same time, it leads to the plasma frequency $\omega_p$ decrease. As the TCO materials are characterized by lower electron density compared to noble metals, their electron heat capacity is much smaller [**19, 20, 29**], thus the electrons in TCOs heat up more and cool down faster compared to noble metals. For the same energy provided to the materials, the electrons in TCO materials heat up to a higher temperature, thus increasing the effective mass. In the absence of light, the relaxation process results in the transfer of energy from the electrons to the lattice in the form of heat.

It is important to note that for a low light frequency and low carrier concentration, the scattering contribution (**Table 1**) can play a significant role as it depends on both the effective mass and the carrier mobility and is expressed by:

$$\gamma = \frac{e}{m^* \mu_{mob}} \quad (4)$$

where $\mu_{mob}$ is the carrier mobility. A longer ENZ wavelength, as in the case of AZO (**Fig. 1a**), corresponds to the lower free carrier density and thus a lower Fermi level. It should be remembered that electrons at lower Fermi level undergo a larger change in effective mass than electrons at higher Fermi level, for the same amount of absorbed energy, resulting in a larger nonlinearity [**17**]. Furthermore, carrier mobility decreases with increasing effective mass [**18**]. However, this relation is not linear and depends on the many contributions, such as the carrier concentration in the TCO and thus the Fermi energy level, the quality of the fabricated TCO thin films, thus the scattering rate can rise or drops under different working conditions. It has been experimentally confirmed that an increase in the effective mass under intraband absorption of light causes the plasma frequency to decrease and the scattering rate to increase at the same time [**18, 30**]. Since both the plasma frequency (**Eq. 3**) and the scattering rate (**Eq. 4**) are inversely proportional to the effective mass, the increase in the scattering rate with an increase in light absorption results in a decrease in mobility.

**Rib photonic waveguide with thin TCO layer**

To investigate a thickness dependence of the TCO materials on the performance of the real devices, a SIN rib photonic waveguide was examined with the TCO thin material, ITO or GZO, placed between the rib and the ridge (**Fig. 2a**), *i.e.*, in the electric field maximum of the propagating mode, which enhances an interaction of light with the TCO material. As shown in **Fig. 2b**, the mode power attenuation, which corresponds to the maximum absorption of light by the waveguide, shifts to lower ENZ plasma frequencies with decreasing ITO thickness. For a constant carrier concentration in the ITO, the decrease in the ENZ plasma frequency can be attributed to the increase in the effective mass in the ITO under illumination of light. Since the calculations were performed at the same power, the increase in the effective mass can be related to the higher electric field confinement in the ITO.

In the next step, the results obtained for a waveguide arrangement with ITO and GZO thin films were compared to show the influence of the plasma frequency on the overall performance, since GZO material exhibit slightly higher plasma frequency compared to ITO as it has been shown in **Table 1**. Higher plasma frequency means higher carrier concentration in GZO. Thus, higher light frequency is required to enter the ENZ regime, which results in lower ENZ wavelength as shown in **Fig. 1a**. Simultaneously, lower permittivity due to bound electrons means that lower plasma frequency is required to reach the ENZ point, and the rib waveguide with GZO layer inside shows maximum attenuation at lower plasma frequencies as shown in **Fig. 2b**.

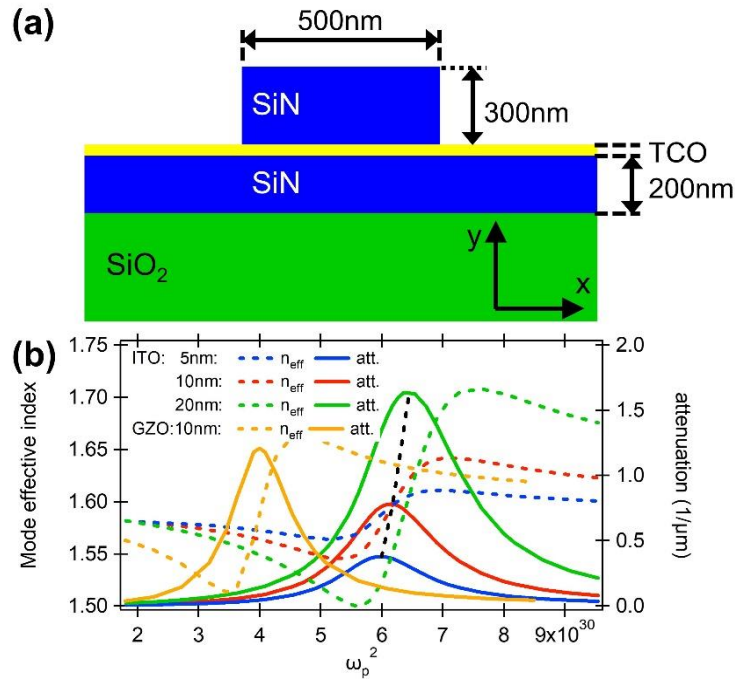

Fig. 2. (a) Geometry of the considered devices and (b) the mode effective index and mode attenuation for different thickness of ITO films and compared with the GZO films.

The obtained results are in good agreement with our previous investigations for thin TCO layers placed in plasmonic [**5, 6**] and photonic [**7**] waveguide arrangement.

**Field enhancement in thin TCO**
For such an arrangement as shown above, the TCO exhibits optical anisotropy and polarization-dependent resonant behavior at ENZ wavelengths. Depending on the polarization of light, *i.e.*, electric field polarized along the width (TE mode) or height (TM mode) of the TCO layer, a different electric field strength can be achieved in the TCO.

The high electric field enhancement in the thin TCO films investigated in this paper under TM polarized light results directly from the boundary conditions, $d_0\varepsilon_0 E_{0\perp} = d_1\varepsilon_1 E_{1\perp}$, which require the phase matching of the total electric fields on both sides of the interface. Here, $d_0$ and $d_1$ are the thicknesses of the waveguide core (before coupling) and the TCO film, $\varepsilon_0$ and $\varepsilon_1$ are permittivities of the waveguide core and the TCO film, respectively, and $E_{0\perp}$ and $E_{1\perp}$ are the normal components of the electric field in the SiN rib waveguide and the TCO, respectively. Thus, for a small $\varepsilon_1$ near the ENZ region, the field normal to the TCO film is strongly enhanced. Furthermore, as the thickness of the TCO is reduced, more energy is provided to the TCO what further enhances the electric field in the TCO and consequently, increases the effective mass. Here, the increase in effective mass is not related to the increases in the carrier concentration as discussed previously [**13, 14, 23**] but rather to the increases in energy provided to the system under an intraband absorption of light. As a result, the carriers are heated and move higher in the conduction band. Due to the non-parabolic density of states, as the electrons move higher in the conduction band their effective mass increases.

**ENZ wavelength evaluation**
In most cases, the increase of the effective mass leads to the redshift of the ENZ wavelength as has been shown in many previously published papers [**4-13**]. However, under some circumstances, the blueshift of the ENZ wavelength can also be expected. It can have a different origin that will be discussed below.

First, as shown previously, the blueshift in the surface plasmon polariton (SPP) mode can be observed for a highly confined SPP mode in a V-groove waveguide structure when the width is reduced from 100 to 5 nm [**31**]. This behavior is expected for structures with either reduced sizes or reduced aspect ratios in the direction of the electric field. Thus, the small thickness of the TCO layer may be one of the mechanisms that can contribute to the blueshift of the ENZ wavelength.

The second mechanism that can contribute to the blueshift of the ENZ wavelength can be related to the effective mass increases under an absorption of light by a thin layer of TCO film. To justify this, the calculations and FEM simulations were performed to show the change of the ENZ permittivity under different effective masses and light frequencies (**Fig. 3 and 4**). The calculations were performed for ITO, for which the carrier mobility is usually measured in the range of 5-30 cm$^2$/(V·s) [**28**].

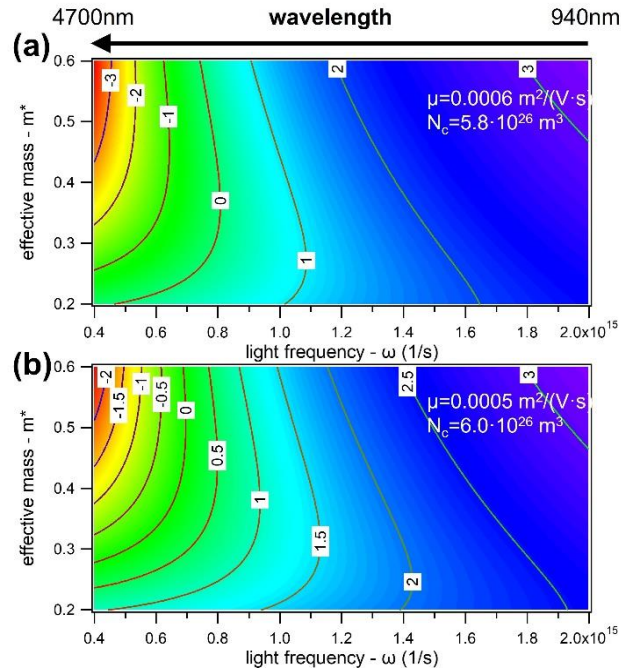

Fig. 3. **Permittivity map (real part)**. (a, b) Real part of a permittivity map in a function of effective mass and light frequency for ITO material with (a) a carrier concentration of 5.8·10$^{26}$ m$^{-3}$ and a carrier mobility $\mu$=6 cm$^2$/(V·s) and (b) a carrier concentration of 6.0·10$^{26}$ m$^{-3}$ and a carrier mobility $\mu$=5 cm$^2$/(V·s).

As shown **Fig. 3**, even a small change in the carrier concentration and carrier mobility is sufficient to change operating conditions of the TCO-based devices. For example, a small decrease in carrier concentration from $N_c$=6.0·10$^{26}$ m$^{-3}$ to $N_c$=5.8·10$^{26}$ m$^{-3}$ with a simultaneous increase in carrier mobility from $\mu$=5 cm$^2$/(V·s) to $\mu$=6 cm$^2$/(V·s) causes a significant shift in the ENZ wavelength from $\lambda_{ENZ}$=2690 nm to $\lambda_{ENZ}$=2350 nm. At the same time, a significant blueshift of the ENZ wavelength corresponding to the $\varepsilon'$=0 line is observed in the MIR wavelengths for a small increase of the initial effective mass. On the contrary, for higher effective masses above 0.35$m^*$ (**Fig. 3a**) and 0.50$m^*$ (**Fig. 3b**), respectively, the expected redshift of the ENZ wavelength is observed. However, since the initial ENZ wavelengths blueshift from $\lambda_{ENZ}$=4090 nm to $\lambda_{ENZ}$=2350 nm is fast for the effective mass increase from 0.20$m^*$ to 0.35$m^*$, the redshift is relatively slow. As the effective mass increase from 0.35$m^*$ to 0.60$m^*$, the ENZ wavelength redshifts from $\lambda_{ENZ}$=2350 nm to $\lambda_{ENZ}$=2540 nm. As we can see from **Fig. 1a**, this is far from the nominal ITO ENZ wavelength of 1500 nm, so a lower carrier concentration in ITO is required to move into the MIR wavelength range.

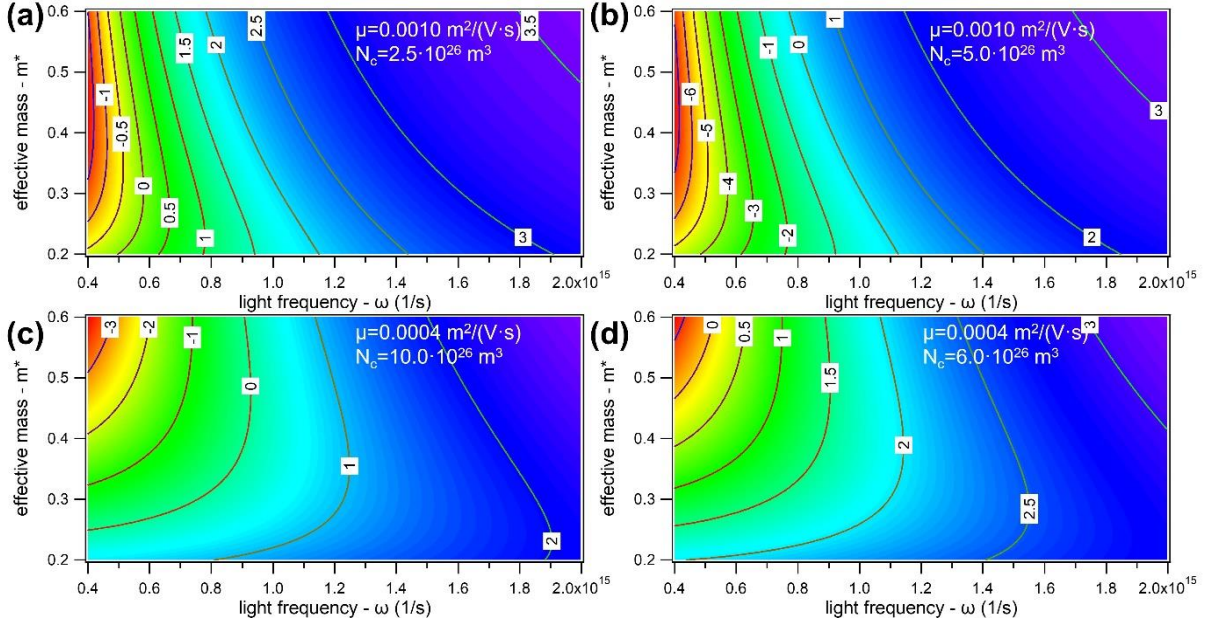

Fig. 4. Real part of a permittivity map in a function of effective mass and light frequency for AZO material with a carrier mobility (a, b) $\mu$=10 cm$^2$/(V·s) and (c, d) $\mu$=4 cm$^2$/(V·s). The carrier mobility was taken at (a) $N_c$=2.5·10$^{26}$ m$^{-3}$, (b) $N_c$=5.0·10$^{26}$ m$^{-3}$, (c) $N_c$=10.0·10$^{26}$ m$^{-3}$, and (d) $N_c$=6.0·10$^{26}$ m$^{-3}$.

Further simulations were performed for a wide range of carrier concentrations and carrier mobilities, showing a huge spectral range of operation under the ENZ conditions (line 0 corresponding to the $\varepsilon'$=0) (**Fig. 4**). For a carrier concentration of 5.0·10$^{26}$ m$^{-3}$ and a carrier mobility of 10 cm$^2$/(V·s), the ENZ wavelength was calculated to be 1670 nm, which is very close to the nominal ENZ value of 1500 nm (**Fig. 1a**). Furthermore, under these operating conditions, the redshift of the ENZ wavelength is observed over the entire range of the effective masses (**Fig. 4b**). This is in good agreement with many other observations and measurements performed for ITO material [**6, 7, 25**]. By decreasing the carrier concentration to $N_c$=2.5·10$^{26}$ m$^{-3}$ while keeping the carrier mobility constant at $\mu$=10 cm$^2$/(V·s), the ENZ wavelength is shifted to the MIR wavelength range (**Fig. 4a**).

To move into the NIR wavelength range, a higher doping of ITO is required as shown in **Fig. 4c**. As the carrier concentration of ITO increases, the carrier mobility decreases as has been shown above. Furthermore, this observation has been confirmed experimentally [**22, 28**] where the mobility decreases from 24 to 9 cm$^2$/(V·s) as the carrier concentration increases. This behavior is expected because increasing the dopant density leads to increased defect scattering, which influences the carrier mobility, that decreases. Therefore, at higher carrier densities, conventional dopants, such as tin in ITO, hybridize with the host conduction band states, increasing the band curvature and hence the effective mass, while decreasing the carrier mobility [**22**]. Here, the calculations were performed for a carrier concentration of 10.0·10$^{26}$ m$^{-3}$ and a carrier mobility of 4 cm$^2$/(V·s), where the ENZ wavelength was calculated to be $\lambda_{ENZ}$=2020 nm, corresponding to a high effective mass of 0.48$m^*$ (**Fig. 4c**). As the carrier concentration increases, the Fermi level moves to the higher level in a conduction band of a nonparabolic material, which has a direct influence on the effective mass that arises even under an absence of the external light source when we assume an electrical doping of the ITO layer [**Sivan**]. Decreasing of the carrier concentration while keeping the carrier mobility at the same level has a huge influence on the ENZ wavelength, which shifts to $\lambda_{ENZ}$=4500 nm for the same effective mass of 0.48$m^*$, as shown in **Fig. 4d**. Such conditions can occur for a low-doped ITO under high light illumination, which causes the effective mass to increase and the corresponding mobility to decrease. As shown above, the TCO materials have great potential to operate efficiently under different wavelength ranges from NIR to MIR. The proper choice of wavelength depends on the properties of the TCO material. TCO has a huge advantage over other materials in that the carrier concentration can be tuned either by interband absorption of light, electrical doping or by incorporating higher levels of dopants into the material during a fabrication process. One of the examples is zinc oxide (ZnO), where doping ZnO with aluminum (Al) or gallium (Ga) increases the carrier concentration that is proportional to the dopant concentration [**14, 24, 25**].

Both ITO and ZnO doped with Al or Ga are characterized by a scattering rate that is one order of magnitude lower than their plasma frequencies, thus the blueshift in the wavelength with an increase in the effective mass can be observed in the NIR and MIR wavelength range. In comparison, indium-doped CdO shows an order of magnitude higher mobility compared to any other TCOs, which strongly influences its scattering rate, *i.e.*, the damping factor, so that the blueshift in the ENZ wavelength can be observed above a wavelength of 10 μm. Below this wavelength, the increase of the effective mass leads to the redshift of the ENZ wavelength as expected by many measurements and observations [**23**].

**Experimental validation from a literature**

The results presented in a recent paper [**25**] confirm the observations presented in this paper about an initial blueshift of the ENZ wavelength under a small effective mass increase. The experiments were performed on a gallium-doped zinc oxide (GZO) under an intraband absorption of light and showed a small blueshift of the ENZ wavelength with a decrease of the GZO layer thickness from 500 nm to about 250 nm. For the GZO material studied, the ENZ wavelength was measured to be 1560 nm. The authors attribute this initial blueshift of the EZN wavelength to the presence of the Fabry-Perot cavity, but the results presented in this article seem to indicate a different basis for this phenomenon. According to the data presented in this paper, as the thickness of the GZO thin film decreases, the electric field in GZO is enhanced, which is a direct result of the boundary conditions, and more energy is provided to the electrons. As a result, the effective mass increases. However, for an initial reduction in the GZO thickness from 500 to 250 nm, the change in effective mass is very small. Further reduction of the GZO thickness below 250 nm reverses this trend and a redshift in the ENZ wavelength can be observed. Below this thickness, even a small change in GZO thickness from 250 nm to 150 nm results in a significant ENZ wavelength redshift from 1400 nm up to 2000 nm as a result of high field enhancement in GZO. It should be noted that for this GZO sample, the carrier concentration was determined to be $N_c$=6.07·10$^{26}$ m$^{-3}$.

*Thickness of TCO vs. carrier concentration*

Furthermore, this article presents a differing view to that presented in Ref. [**23**] and subsequently reproduced in Ref. [**13**] and Ref. [**14**], namely that an increase in the thickness of the TCO film leads to an increase in the carrier concentration - there is as yet no evidence to support this hypothesis. Even the data presented in the above article [**23**] seems to contradict these statements of the authors where for different ITO thicknesses of 130, 200, 306 and 1060 nm the carrier concentration was measured at 3.6·10$^{20}$, 5.1·10$^{20}$, 4.9·10$^{20}$, and 4.1·10$^{20}$ cm$^{-3}$, respectively. Thus, it is not possible to say that the carrier concentration depends on the ITO thickness.

On the other hand, it can be observed that only the mobility shows some consistency and increases regularly with increasing ITO thickness from 15 to 21.8 cm$^2$/(V·s). As shown in many papers previous works on TCO material [**22, 28**], the mobility decreases with increasing carrier concentration due to the nonparabolicity of the conduction band. Therefore, the second factor influencing the plasma frequency should be considered namely, the effective mass. As the thickness of the TCO thin film decreases, more energy is provided to the electrons in the TCO material, what shifts the Fermi energy to a higher level in the conduction band. Due to the nonparabolicity of the conduction band, the effective mass of the electron sea increases. Therefore, the "heavier" electrons are characterized by a lower mobility. In Ref. [**24**] the authors mention that thinner TCOs result in lower conductivity, but this is not necessarily related to lower carrier concentration. In reality, the electrical conductivity depends on both the carrier concentration and the carrier mobility according to the equation $\sigma=ne\mu_{mob}$, so the lower conductivity for thinner TCO materials can be explained by the lower mobility as a result of the effective mass increase for thinner TCO materials.

**Conclusion**

Here we have shown that placing the thin TCO film in the maximum electric field of the propagating mode can provide a significant modulation potential, which is highly dependent on light polarization. Due to high field confinement in TCO, electron temperature in TCO increases significantly, resulting in effective mass increase. In most cases, this causes the redshift of the ENZ wavelength. However, under certain conditions, such as proper carrier concentration and carrier mobility in the TCO, excitation wavelength, and light power, a blueshift in the ENZ wavelength may occur. Furthermore, the operating conditions such as the operating wavelength can be adjusted by a proper choice of TCO material and initial doping level. Thus, tuning the ENZ wavelength over a broad spectrum range from UV to MIR can offer significant advantages for active components that exclusively operate under optical switching mechanisms. However, further studies are required to explain in depth the mechanism behind this process.

**Acknowledgements**

The author acknowledges the constant support of Warsaw University of Technology, Poland for the completion of this work. Furthermore, J.G. is very thankful to Prof. D. G. Misiek for his support and very valuable suggestions.